\documentclass[a4paper,fleqn,usenatbib]{mnras}

\usepackage{newtxtext,newtxmath}

\usepackage[T1]{fontenc}
\usepackage{ae,aecompl}

\usepackage{graphicx}	
\graphicspath{{Figures/}}
\usepackage{amsmath}	
\usepackage{amssymb}	
\usepackage{color}
\usepackage{subfigure}

\title[Characterising relative orientations]{An application of an optimal statistic for characterising relative orientations}

\author[D. L. Jow et al.]{
Dylan L. Jow,$^{1}$\thanks{E-mail: d.jow@alumni.ubc.ca}
Ryley Hill,$^{1}$
Douglas Scott,$^{1}$
J.D. Soler,$^{2}$
P.G. Martin,$^{3}$
\newauthor
\ M.J. Devlin,$^{4}$
L.M. Fissel$^{5}$
and F. Poidevin$^{6,7}$
\\
$^{1}$Department of Physics \& Astronomy, University of British Columbia, 6224 Agricultural Road, Vancouver, BC V6T 1Z1\\
$^{2}$Max-Planck-Institute for Astronomy, K\"onigstuhl 17, 19117, Heidelberg, Germany \\
$^{3}$CITA, University of Toronto, Toronto, ON M5S 3H8, Canada \\
$^{4}$Department of Physics \& Astronomy, University of Pennsylvania, Philadelphia, PA 19104, U.S.A. \\
$^{5}$National Radio Astronomy Observatory, 520  Edgemont Rd, Charlottesville, VA 22903, U.S.A. \\
$^{6}$Instituto de Astrofsica de Canarias, E-38205 La Laguna, Tenerife, Spain. \\
$^{7}$Universidad de La Laguna, Dpto. Astrofsica, E-38206 La Laguna, Tenerife, Spain.
}

\date{Accepted XXX. Received YYY; in original form ZZZ}

\pubyear{2017}

\begin{document}
\label{firstpage}
\pagerange{\pageref{firstpage}--\pageref{lastpage}}
\maketitle

\begin{abstract}
We present the projected Rayleigh statistic (PRS), a modification of the classic Rayleigh statistic, as a test for non-uniform relative orientation between two pseudo-vector fields. In the application here this gives an effective way of investigating whether polarization pseudo-vectors (spin-2 quantities) are preferentially parallel or perpendicular to filaments in the interstellar medium. For example, there are other potential applications in astrophysics, e.g., when comparing small-scale orientations with larger-scale shear patterns. We compare the efficiency of the PRS against histogram binning methods that have previously been used for characterising the relative orientations of gas column density structures with the magnetic field projected on the plane of the sky. We examine data for the Vela C molecular cloud, where the column density is inferred from \textit{Herschel} submillimetre observations, and the magnetic field from observations by the Balloon-borne Large-Aperture Submillimetre Telescope in the 250-, 350-, and 500-$\mu$m wavelength bands. We find that the PRS has greater statistical power than approaches that bin the relative orientation angles, as it makes more efficient use of the information contained in the data. In particular, the use of the PRS to test for preferential alignment results in a higher statistical significance, in each of the four Vela C regions, with the greatest increase being by a factor 1.3 in the South-Nest region in the 250-\,$\mu$m band. 
\end{abstract}

\begin{keywords}
polarization -- methods: statistical -- ISM: dust, magnetic fields, clouds 
\end{keywords}



\section{Introduction}

Directional data, i.e. quantities that have an orientation, arise frequently and across many disciplines, from the study of bird migration patterns \citep[e.g.,][]{Cochran405} to meteorology and the direction of wind and wave currents \citep[e.g.,][]{BOWERS200013}. The diversity of directional statistics is evidenced by its direct application in military science \citep[e.g.,][]{mahan1991circular} and handwriting analysis \citep[e.g.,][]{Bahlmann:2006:DFO:1220964.1221165}. A common question of interest is the characterisation of the relative orientations between two sets of overlapping directional data, i.e. determining whether or not there is statistical alignment, and more specifically whether or not the two sets of directions are preferentially parallel or perpendicular to each other. A question of this form was posed by \citet{Burda07042009}, where it was determined that herding cows tend to align with the Earth's magnetic field. The subsequent criticism of this finding by \citet{Hert2011} and rebuttal of that criticism by \citet{Begall2011} illustrates the need for robust statistical modelling of orientation effects. From the point of view of the astronomer, however, the subject of cow alignment is moot.

Magnetic fields are important in astrophysics, but cows are replaced by molecular clouds, in the formation of which magnetic fields are believed to play an important role \citep[e.g.,][]{2012ARA&A..50...29C,2012SSRv..166..293H}. Recently, particular interest has been given to examining the relationship between the orientations of the magnetic field projected in the plane of the sky, $\langle\hat{\bf{B}}_{\perp}\rangle$, and the morphology of structures in the total column density of the interstellar medium. \cite{2013ApJ...774..128S} describe the histogram of relative orientations (HRO) method, which characterises the column density orientation by its gradient, and the magnetic field through polarization data. Polarization is characterised by pseudo-vectors, which are vector-like quantities that are invariant under 180 degree rotations. Formally, they are spin-2 quantities. The alignment of the pseudo-vector fields is then analysed through the use of a shape parameter (the HRO statistic) defined by the differences of areas computed from the histogram of relative angles. This method was used in \citet{2016A&A...586A.135P} to determine that the $\langle\hat{\bf{B}}_{\perp}\rangle$ field inferred from the \textit{Planck} 353-GHz (850-$\mu$m) polarization data at 10\,arcmin resolution are predominantly parallel to filamentary structures in the diffuse interstellar medium (ISM) contours. Furthermore, \citet{2016A&A...586A.138P} determined that within ten nearby ($d <$ 450\,pc) Gould Belt molecular clouds, the relative orientations of the column density, $N_\mathrm{H}$ inferred from the \textit{Planck} dust emission maps, and the $\langle\hat{\bf{B}}_{\perp}\rangle$ inferred from the \textit{Planck} 353-GHz polarization maps at 10\,arcmin resolution changed progressively from preferentially parallel to perpendicular with increasing $N_\mathrm{H}$. \cite{2017A&A...603A..64S} examined the relationship between the magnetic field and column density morphology by applying the HRO statistic to the magnetic field inferred from the Balloon-borne Large-Aperture Submillimetre Telescope for Polarimetry (BLASTPol) observations in three different wavelength-bands (250, 350, and 500\,$\mu$m) for the Vela C molecular cloud. 

In this paper, we present an alternative to the HRO statistic, which we call the projected Rayleigh statistic (PRS). As with the HRO statistic, the PRS can be used to test for preferentially parallel or perpendicular alignment, in a modification of the classic Rayleigh test widely used in circular statistics \citep[see e.g.,][]{batschelet1981circular,BIMJ:BIMJ4710380307,Mardia-directional-stats}. We compare the two approaches using simulated models, and for the Vela C region using the same BLASTPol and \textit{Herschel} observations, as described in \cite{2017A&A...603A..64S}. From our results, we argue that the PRS is the optimal statistic for this type of analysis, although with open issues regarding the choice of weighting scheme. The PRS can also be applied to the investigation of alignment between small and large-scale orientations, for example, investigating individual galaxy or binary system orientation within larger elliptical structures or clusters. Therefore, we suggest that the PRS has the potential for broad application in astronomy.

\section{Observations}
\label{sec:observations}

In this work, we use two data sets: the Stokes $I$, $Q$ and $U$ observations obtained during the 2012 flight of BLASTPol \citep{2014SPIE.9145E..0RG}; and the total column density maps derived from the \textit{Herschel} satellite dust-continuum observations \citep{2010A&amp;A...518L...1P}. We specifically use the BLASTPol polarization and \textit{Herschel} column density maps of the Vela C molecular cloud already presented in \citet{2017A&A...603A..64S}, where these objects are described in detail. The diffuse Galactic emission contribution to the polarization maps was subtracted according to the intermediate method described in \cite{2016ApJ...824..134F}. 

The magnetic field pseudo-vectors, $\langle\hat{\bf{B}}_{\perp}\rangle_\lambda$, for each polarization band (characterised by its central wavelength $\lambda$) were inferred from the polarization angle given by

\begin{equation}
	\psi_\lambda = \frac{1}{2}\arctan(-U_\lambda,Q_\lambda) \text{\footnotemark}.
	\label{eq:pol_phi}
\end{equation}
\footnotetext{Note that here we are using the International Astronomical Union convention.}

The polarization angle determines the direction of the polarization pseudo-vector, $\hat{\bf{P}}_\lambda$, and we infer the magnetic field orientation by assuming that it is perpendicular to the polarization field \citep{1988QJRAS..29..327H}. The gradient map of the column density was obtained using the Gaussian derivatives method described in \citet{2013ApJ...774..128S}, convolving with a $5\times5$ Sobel kernel. 

\subsection{The Vela C region}
\label{sec:Vela_C_region}

\begin{figure}
	\includegraphics[width=\columnwidth]{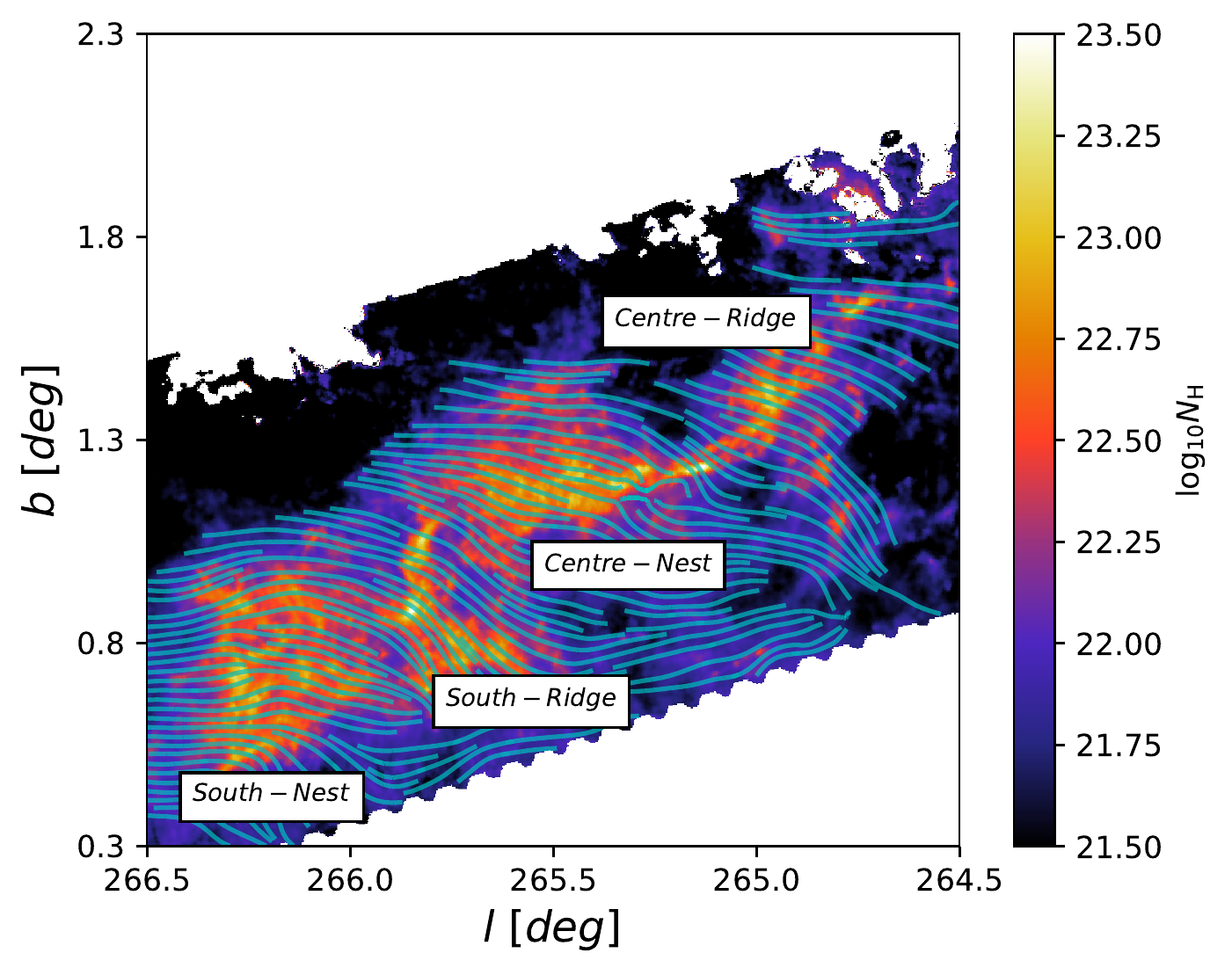}
	\caption{Column density of the Vela C molecular cloud, inferred from the \textit{Herschel} observations, with labelled sub-regions as defined by \citet{2011A&amp;A...533A..94H}. The faded cyan lines indicate the magnetic field projected on the plane of the sky, taken to be orthogonal to the BLASTPol 500-$\mu$m polarization pseudo-vectors. The column density, $N_\mathrm{H}$, is reported in units of $\mathrm{cm}^{-2}$. The lines were generated using Matplotlib.pyplot's {\tt streamplot} function \citep{Hunter:2007}.}
	\label{fig:VelaC}
\end{figure}

The BLASTPol and column density maps used in this study are of the Vela C molecular cloud, one of four sub-regions of the Vela Molecular Ridge, which lie in the Galactic plane at distances of approximately 700\,pc to 2\,kpc \citep{1991A&A...247..202M,1992A&A...265..577L}. The total molecular mass of the Vela Molecular Ridge is approximately $5 \times 10^5\,\mathrm{M}_\odot$ \citep{1988A&AS...73...51M,1999PASJ...51..775Y,2009ApJ...707.1824N}. The Vela C region in particular is considered to be a rare example of a nearby and massive cloud in an early evolutionary stage \citep{2004ApJ...614..818B,2009ApJ...707.1824N,2017A&A...603A..64S}. \citet{2011A&amp;A...533A..94H} describes five further subdivisions of the Vela C cloud, each having distinct characteristics, which they name the North, Centre-Ridge, Centre-Nest, South-Ridge, and South-Nest. The overlap of the BLASTPol and column density maps used in this study contain the latter four of these regions. Figure~\ref{fig:VelaC} shows the $\textit{Herschel}$-derived column density map of Vela C, with the four sub-regions labelled and the magnetic field derived from the BLASTPol data superimposed.

\section{Characterising Relative Orientations}
\label{sec:math_description}

We would like to find an efficient (and, ideally, optimal) statistic with which to characterise the relative alignment of the magnetic field and column density structures. The question we are adressing in its most general form can be stated as follows: given two vector fields $\bf{U}$ and $\bf{V}$, are their relative orientations entirely random, or do they tend to align in some fashion? For example, $\bf{U}$ might encode the orientation of all the cows on Earth, and $\bf{V}$ might be the Earth's magnetic field, and the question would be whether cows' orientations are random with respect to the magnetic field. To answer the problem, one would first calculate the relative angles between each vector in $\bf{U}$ with respect to the vector in $\bf{V}$ at the same position. This would yield a set of angles with a range from $[0,2\pi)$, and the question is then equivalent to determining how these angles are distributed in that range. No preference in orientation would yield a statistically uniform distribution. 

We further restrict our attention to cases where we are interested only in the alignment of the vectors, and not the parity of the alignment. In the example of the cows, we might want to first examine their relative orientation with the Earth's magnetic field, not whether their heads tend to point North or South. In these cases, any relative angle $\phi$ is equivalent to $\phi + \pi$. This can be achieved by first taking the tangent of the relative angles, and then taking the inverse tangent, since $\tan\phi = \tan(\phi+\pi)$. So for any vector $\bf{u}\in\bf{U}$ and the corresponding $\bf{v}\in\bf{V}$, we can calculate their relative angles according to 

\begin{equation}
	\phi = \arctan{\frac{\lvert \bf{u} \times \bf{v} \rvert}{\bf{u} \cdot \bf{v}}}.
	\label{eq:relative_angle}
\end{equation}
This yields a set of angles $\{\phi_i\}$ with range $[-\frac{\pi}{2},\frac{\pi}{2}]$. Note that, as implemented, the norm in Eq.~\ref{eq:relative_angle} carries a sign.

To decide whether two sets of orientations have a tendency to be either parallel or perpendicular to each other, we introduce the histogram of relative orientations method first presented by \citet{2013ApJ...774..128S}, as well as describing a new statistical method, which we call the projected Rayleigh statistic.

\subsection{The histogram of relative orientations}
\label{sec:HROs}

An obvious way to investigate the underlying distribution of a sample of angles is to plot a histogram of that sample. \cite{2013ApJ...774..128S} and \cite{2017A&A...603A..64S} describe what they call the HRO approach, a technique for analysing the relative orientations of interstellar magnetic fields and emission morphologies in molecular clouds. As shown by simulations of magneto-hydrodynamic turbulence, the relative orientation between the magnetic field and column density structures is related to the magnetization in a molecular cloud \citep{2013ApJ...774..128S}. The HRO technique involves characterising the magnetic field direction via sky polarization data, and the molecular cloud orientation through the gradient of the column density and then calculating their relative orientation angles. The analysis of the resulting set of relative angles is carried out by plotting a histogram of angles, the HRO, and analyzing its shape. In particular, \cite{2013ApJ...774..128S} define the HRO ``shape parameter" of a set of angles $\{\phi_i\}$ with range $[-\frac{\pi}{2},\frac{\pi}{2}]$ as $\zeta \equiv A_\mathrm{c} - A_\mathrm{e}$, where $A_\mathrm{c}$ is the area of the central region of the histogram ($-22.5^\circ < \phi < 22.5^\circ$) and $A_\mathrm{e}$ is the area of the edge regions ($-90^\circ<\phi<-67.5^\circ$ and $67.5^\circ<\phi<90^\circ$). The shape parameter is then defined by normalizing by the total area, and also restricting the range of $\{\phi_i\}$ to be $[0,\frac{\pi}{2}]$, mapping all angles to their absolute values. 

As in \cite{2017A&A...603A..64S} we normalize the parameter, but allow the angles to take the full range $[-\frac{\pi}{2},\frac{\pi}{2}]$; this does not imply any loss of generality, since we are unconcerned with the sign of the alignment, so $\phi$ is equivalent to $-\phi$. Thus, the shape parameter is defined as

\begin{equation}
	\zeta \equiv \frac{A_\mathrm{c}-A_\mathrm{e}}{A_\mathrm{c}+A_\mathrm{e}}.
	\label{eq:zeta}
\end{equation}
A value of $\zeta > 0$ implies a preference towards parallel alignment, whereas $\zeta<0$ implies perpendicular alignment. In the case of random alignment, we expect $\zeta$ to be close to $0$, with an uncertainty we can calculate. The uncertainty in $\zeta$, $\sigma_\zeta$, is given by

\begin{equation}
	\sigma^2_\zeta = \frac{4(A^2_\mathrm{e}\sigma^2_{A_\mathrm{c}} + A^2_c\sigma^2_{A_\mathrm{e}})}{(A_\mathrm{c}+A_\mathrm{e})^4},
	\label{eq:dzeta}
\end{equation}
as derived in \citet{2017A&A...603A..64S} .

While simple in form and execution, this shape parameter cannot be the optimal statistic for characterising relative orientations. For one thing, it entirely ignores angles in the ranges of $-67.5^\circ<\phi<-22.5^\circ$ and $22.5^\circ<\phi<67.5^\circ$, and so will be insensitive to large parts of the data. In particular, it will be completely blind to even very strong $45^\circ$ preferential alignment. This problem can, of course, be resolved by simply changing the selected areas since the choice of angle ranges is arbitrary, but any HRO-like statistic must suffer from the intrinsic deficiencies of binning, namely, that binning weights all data in the same bin equally, and so fails to utilise the full power of the data. We suggest, therefore, that a more optimal statistic for analyzing relative orientations would not involve any binning of the angles, and next we describe just such an approach.

\subsection{The projected Rayleigh statistic}
\label{sec:Rayleigh_math} 

Given a set $\{\theta_i\}$ of $n$ angles, with a range of $[0,2\pi]$, to determine whether or not the angles are uniformly distributed, one might use the Rayleigh test, which uses the Rayleigh statistic defined as

\begin{equation}
	Z = \frac{(\sum_{i}^{n}\cos\theta_{i})^2+(\sum_{i}^{n}\sin\theta_{i})^2}{n},
	\label{eq:Rayleigh_stat}
\end{equation}
\citep[see e.g.,][]{batschelet1981circular,BIMJ:BIMJ4710380307,Mardia-directional-stats}.

The Rayleigh statistic is related to a random walk, characterising the distance from the origin if one were to take unit steps in the direction determined by each angle. The expectation value of a random walk in two dimensions is zero, and so any significant deviation from the origin would indicate some sort of preference in the angles. The Rayleigh statistic can be used to test for the non-uniformity of a set of angles, and has already been applied many times in astrophysics when testing for periodicity \cite[e.g.,][]{1982Natur.296..833G,1991ASPC...21..264S,2011APh....34..627P} or modulation of data \cite[e.g., section 6.6 of][]{2016A&A...594A..16P}.

For our purposes, we begin with a set of axial data, $\{\phi_i\}$, with the range $[-\frac{\pi}{2},\frac{\pi}{2}]$. To test against uniformity of these data, we can first convert the axial data to angular data by mapping each angle, $\phi_i$, to twice itself, i.e. $\phi_i \to \theta_i = 2\phi_i$. This method of angle doubling is a common technique for converting axial data to circular data, in order to utilise the rich field of circular statistics \citep{batschelet1981circular, Mardia-directional-stats}. Having doubled our angles, we can simply apply the Rayleigh statistic defined above to test against uniformity, as did \cite{Burda07042009} in their analysis of the relative orientations of cows with magnetic fields. However, for the astrophysical applications we have in mind, we are interested in something slightly more specific than a test against uniformity. In particular, we are interested in whether alignment tends to be preferentially parallel (corresponding to $\phi = 0$), or preferentially perpendicular ($\phi = -\frac{\pi}{2}$ or $\frac{\pi}{2}$), as well as the statistical strength of that alignment. When we multiply our axial data by 2, parallel orientation corresponds to $\theta=0$ while perpendicular orientation corresponds to $\theta=\pi$. Thus, with $\theta=0$ corresponding to the $x$-axis, the quantity of interest is the horizontal distance moved in our hypothetical random walk. 

\citet{10.2307/30080924} describe what they call the \textit{V} statistic, which computes just this quantity. The \textit{V} test is also discussed, under a different name, by \citet{Mardia-directional-stats}. In its most general form, the \textit{V} statistic is used in tests for uniformity against an alternative with specified mean direction. Here, we take the specified direction to be $\theta=0$, and we rename the \textit{V} statistic to the more intuitive ``projected Rayleigh statistic (PRS)". Then, for a set of $n$ angles $\{\theta_i\}=\{2\phi_i\}$ where $\phi_i \in [-\frac{\pi}{2},\frac{\pi}{2}]$, the PRS is
\begin{equation}
	Z_x = \frac{\sum_{i}^{n}\cos\theta_i}{\sqrt{n/2}}.
	\label{eq:projected_Rayleigh}
\end{equation}
Thus, $Z_x \gg 0$ indicates strong parallel alignment, while $Z_x \ll 0$ indicates strong perpendicular alignment. 

As a test of the statistical significance of relative alignment, we can investigate the probability distribution of the PRS when each $\phi_i$ (or equivalently $\theta_i$) is uniformly distributed. Assuming that the angles are independently and uniformly distributed, it follows that every $\cos\theta_i$ is independently and identically distributed, with a mean of $0$. Also, for uniformly distributed $\theta_i$, $\cos^2\theta_i$ has a mean of $\frac{1}{2}$. Therefore, each $\cos\theta_i$ is independently and identically distributed with mean $\mu=0$ and $\sigma^2=\frac{1}{2}$. By the central limit theorem, it follows that $\sum_{i}^{n}\cos\theta_i/\sqrt{n}$ is distributed as $\mathcal{N}(0,\frac{1}{2})$ in the limit that $n\to\infty$, where $\mathcal{N}(\mu,\sigma^2)$ is the normal (or Gaussian) distribution with mean $\mu$ and variance $\sigma^2$. 

The asymptotic limit of the PRS distribution is, therefore, the standard normal distribution. For a general distribution of angles the variance in $Z_x$ (in the high-$n$ limit) is simply the variance of each $\cos\theta_i/\sqrt{1/2}$, which can be estimated as

\begin{equation}
	\sigma_{Z_x}^2 = \frac{2\sum_{i}^{n}(\cos\theta_i)^2 - (Z_x)^2}{n}.
	\label{eq:Zx_variance}
\end{equation}
We take this value to be the uncertainty in the PRS. That is, the uncertainty estimated for a single PRS measurement reflects the dispersion of the relative angles.

The Rayleigh test is equivalent to the likelihood ratio test against von Mises alternatives (see Section~\ref{sec:statistical_power} for a description of the von Mises distribution), the likelihood ratio test being optimal by the Neyman-Pearson lemma \citep{Mardia-directional-stats}. Therefore, the PRS is necessarily more powerful than the HRO shape parameter, and indeed any other test, when testing against alternatives to von Mises distributions with known mean. The optimality of the Rayleigh statistic in this case is also discussed in \citet{10.2307/30080924}. It should be noted that if the shape of the underlying distribution of angles is known, then one could always find a better test for that specific distribution. However, as a test for orientation in general, the PRS is clearly optimal.

An additional advantage of using the PRS is the ability to implement a weighting scheme. Whereas binning methods may need to reject noisy data, the PRS can still utilise these data by weighting them accordingly. Thus, the PRS can be modified to use the full power of the data, including the noisy parts. The weighted projected Rayleigh statistic can be defined as

\begin{equation}
	Z^{*}_{x} = \frac{\sum_{i}^{n}w_i\cos\theta_i}{\sqrt{\sum_{i}^{n}(w_i)^2/2}},
	\label{eq:Zx_weighted}
\end{equation}
where $w_i$ is the weight for the angle $\theta_i$. The weighted PRS for uniformly distributed relative angles follows the same distribution as the PRS, namely, the standard normal distribution. In general, the weights will be related to the noise in individual data samples, as we discuss in Sect.~\ref{sec:polarization} for a particular example. The form of the weights can be tuned for each specific application.

\subsection{The PRS and HRO distributions for uniform relative angles}
\label{sec:pdf_model}

In order to utilise the HRO and PRS methods as tests of orientation in real data we must have a description of their probability distributions under the conditions of purely random relative orientations. In Section~\ref{sec:Rayleigh_math} we derived the asymptotic form of the PRS as the number of angles $n$ approaches infinity. However, we have not determined the size at which $n$ is ``large enough", and, in any case, we should check the results with numerical simulations.

\begin{figure}
\subfigure{\includegraphics[width=\columnwidth, height=5.5cm]{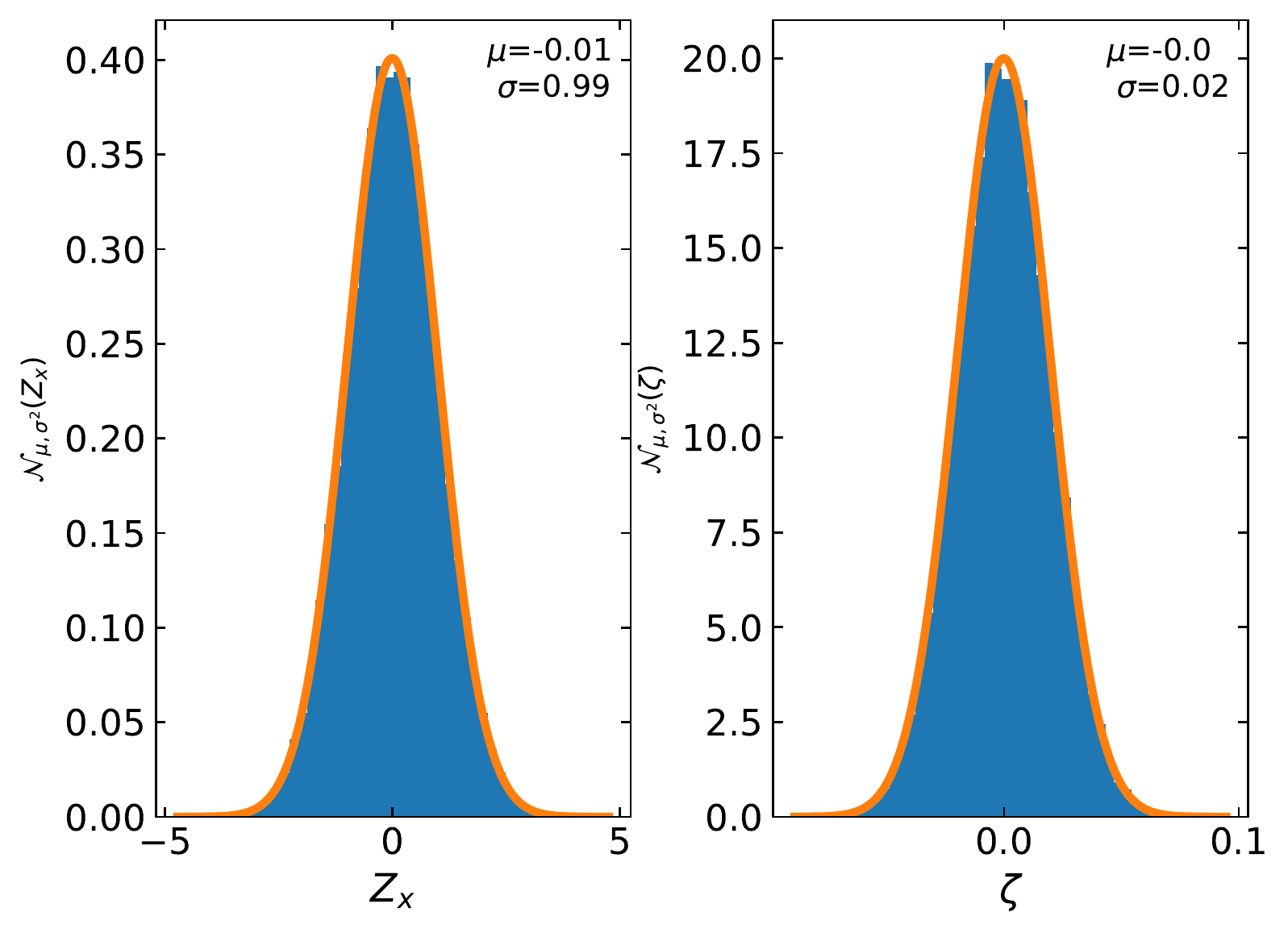}}
\subfigure{\includegraphics[width=\columnwidth, height=5.5cm]{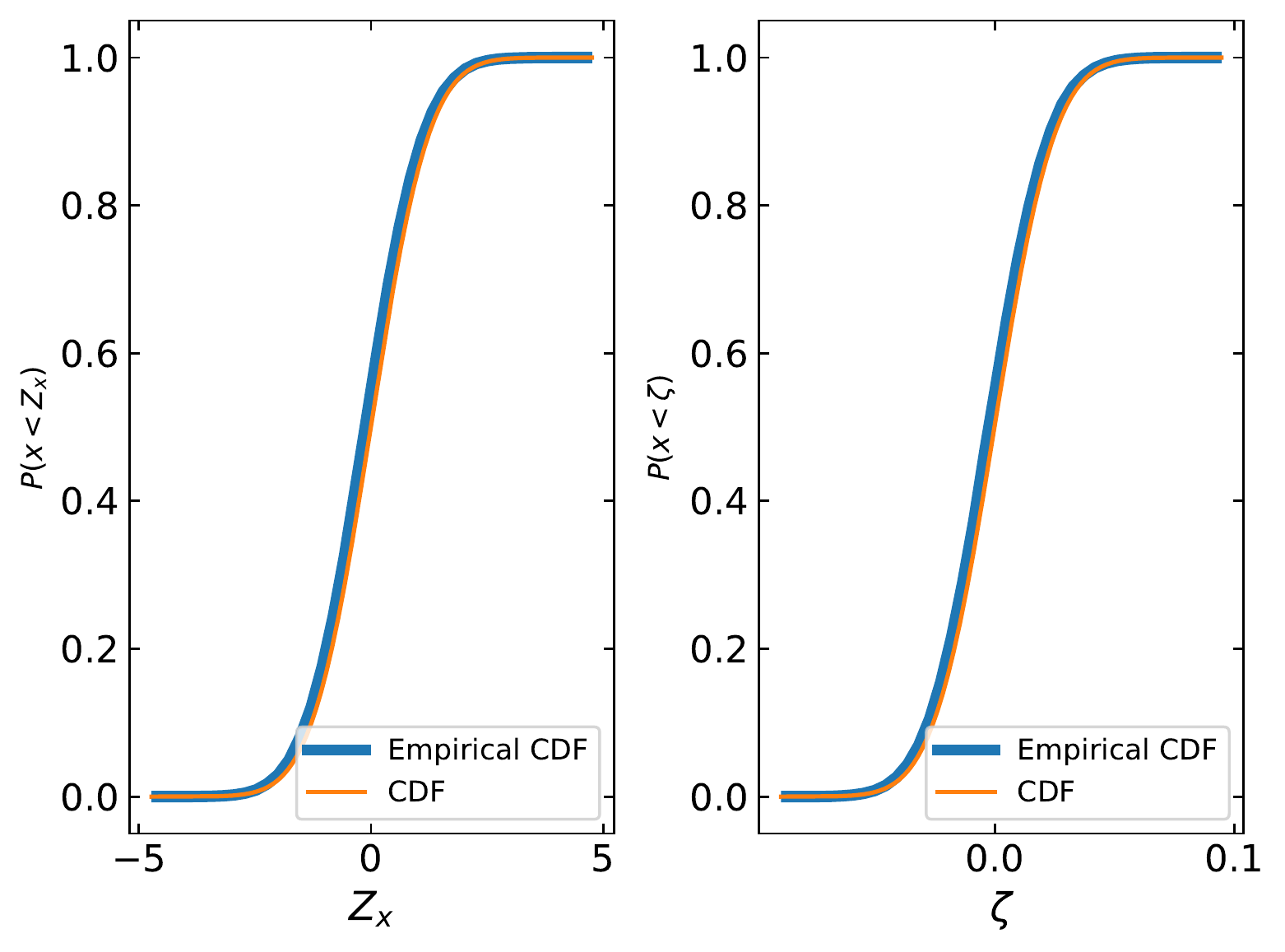}}
	\caption{Top: Empirical PDFs of the PRS, $Z_x$ (left) and HRO shape parameter, $\zeta$ (right).  These are derived from 100,000 evaluations of the statistic on samples of angles ($n = 5000$) drawn from a uniform distribution.  They are fit with a Gaussian for comparison to the exact normal PDFs expected for high $n$. Bottom: The respective empirical CDFs, compared to those expected for high $n$.} 
	\label{fig:Z_Zeta_pdfcdf}
\end{figure}

We compute the PRS and HRO shape parameter for $N=100{,}000$ times for different samples of $n$ angles drawn from a population uniformly distributed on $[-\frac{\pi}{2},\frac{\pi}{2}]$. We use this to determine the empirical probability distribution functions (PDFs). The top panel of Fig~\ref{fig:Z_Zeta_pdfcdf} shows the distributions of the PRS and the HRO shape parameter for $n=5000$, fitted to an exactly normal distribution. This verifies that the PRS is distributed according to the standard normal distribution, and that the HRO shape parameter is also normally distributed. Since for statistical tests, we are typically more concerned with the tails of the distribution we also compare the empirical cumulative distribution functions (CDFs) with the exactly normal CDFs and find very good agreement (bottom panel of Fig.~\ref{fig:Z_Zeta_pdfcdf}). We find that $n>5000$ yields 0.01 as an upper bound on the difference between the empirical and theoretical CDFs for both the PRS and HRO shape parameter. Applying the Kolmogorov-Smirnov test, we find that this bound would occur with 70 per cent frequency given the hypothesis that the two distributions are the same. Thus, there is no significant difference between the empirical distributions and the hypothesized normal distributions. In the analyses that follow, we use samples of relative angles with size $n>5000$ to ensure that the distributions of the PRS and HRO shape parameter are normal. For smaller sample sizes, the PDF histograms could still be used, but the statistical significance would need to be computed using numerically generated PDFs of both statistics, rather than just assuming Gaussian distributions.

It is important to note that while $Z_x$ is distributed according to $\mathcal{N}(0,1)$ for any sufficiently large number of angles, $n$, we find that $\zeta$ is distributed according to $\mathcal{N}(0,\sigma_\zeta^2(n))$, where the variance $\sigma^2_\zeta(n)$ is dependent on $n$. Therefore, in the analyses that follow, when we determine the statistical significance of $\zeta$ we recompute and fit the probability distribution of $\zeta$ for the appropriate $n$ to determine the variance.

\subsection{Statistical power}
\label{sec:statistical_power}

The power of a test statistic is defined as the probability that it rejects the null hypotheses given that the alternative hypothesis is true. We are interested in examining and comparing the statistical power of the PRS, $Z_x$, and the HRO shape parameter, $\zeta$. To do so, we take $N$ samples of size $n$ from a population distributed non-uniformly on $[-\frac{\pi}{2},\frac{\pi}{2}]$. For each sample, we calculate both $Z_x$ and $\zeta$ and define a detection to be a value with statistical significance greater than a threshold, say $>5\sigma$. Then the statistical power is simply the fraction of the $N$ samples in which that statistic is deemed a detection.

Here we verify numerically that the PRS is more powerful than the HRO shape parameter as a test against von Mises alternatives. The von Mises distribution is a probability distribution of a random variable in the range of $[-\pi,\pi)$. It is a close approximation of the wrapped normal distribution, and can be thought of as a circular analogue of a Gaussian distribution \citep{BIMJ:BIMJ4710380307,Mardia-directional-stats}. It is defined as

\begin{equation}
	\mathcal{F}(\theta \vert \mu,\kappa) \equiv \frac{e^{\kappa\cos{\theta-\mu}}}{2\pi I_0(\kappa)},
	\label{eq:vonMises}
\end{equation}
where $I_0(\kappa)$ is the modified Bessel function of order 0. Here, the parameter $\mu$ is the mean of the distribution, and $1/\kappa$ characterises the dispersion, which is analogous to $\sigma^2$ for the normal distribution. As $\kappa$ decreases to 0, the distribution continuously deforms to the uniform distribution on $[-\pi,\pi)$ \citep{Mardia-directional-stats}. We are interested in a population distributed on $[-\frac{\pi}{2},\frac{\pi}{2}]$, rather than the full circle, so we simply divide all of our sample angles by 2.

\begin{figure}
	\subfigure{\includegraphics[width=\columnwidth]{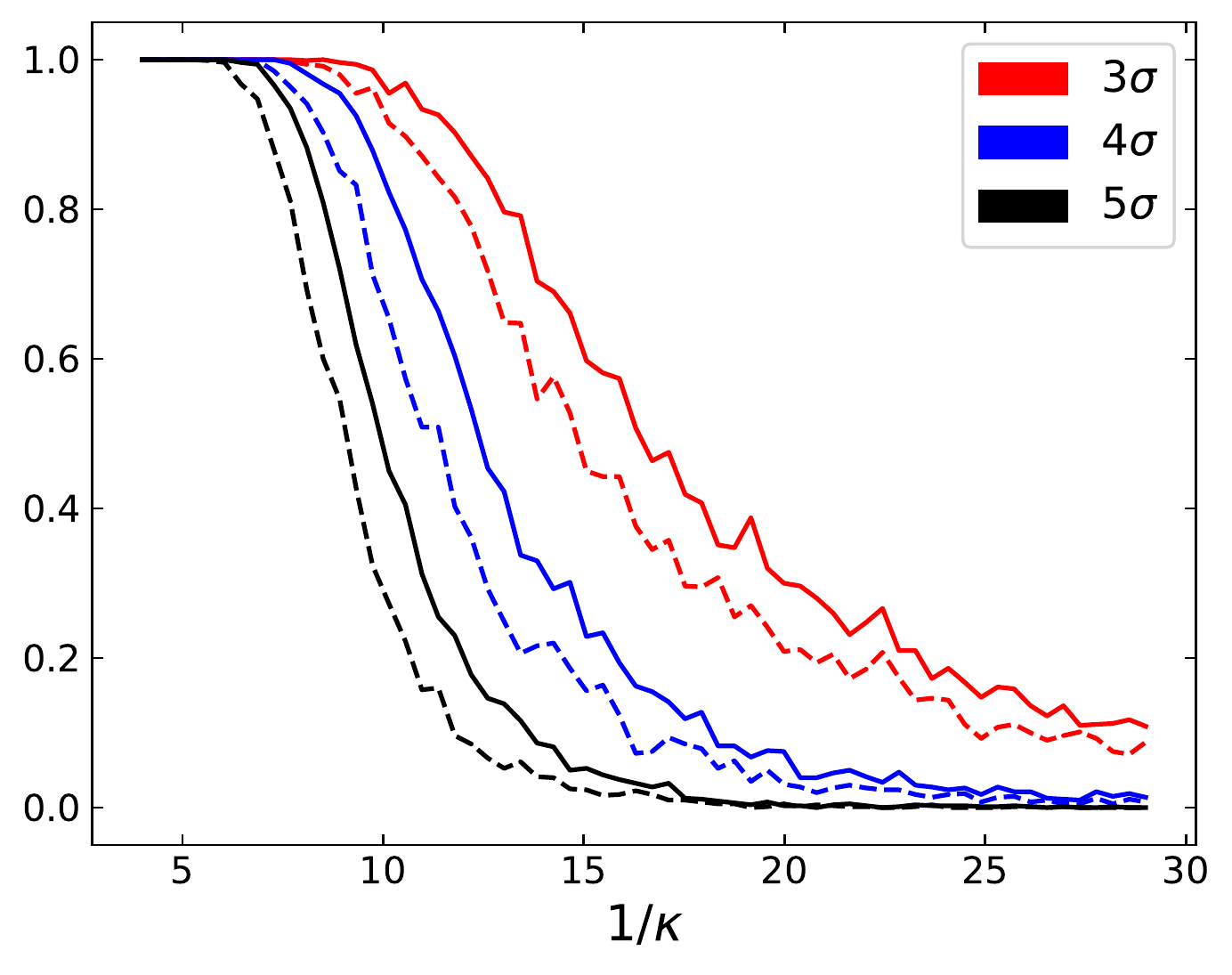}}
	\subfigure{\includegraphics[width=\columnwidth]{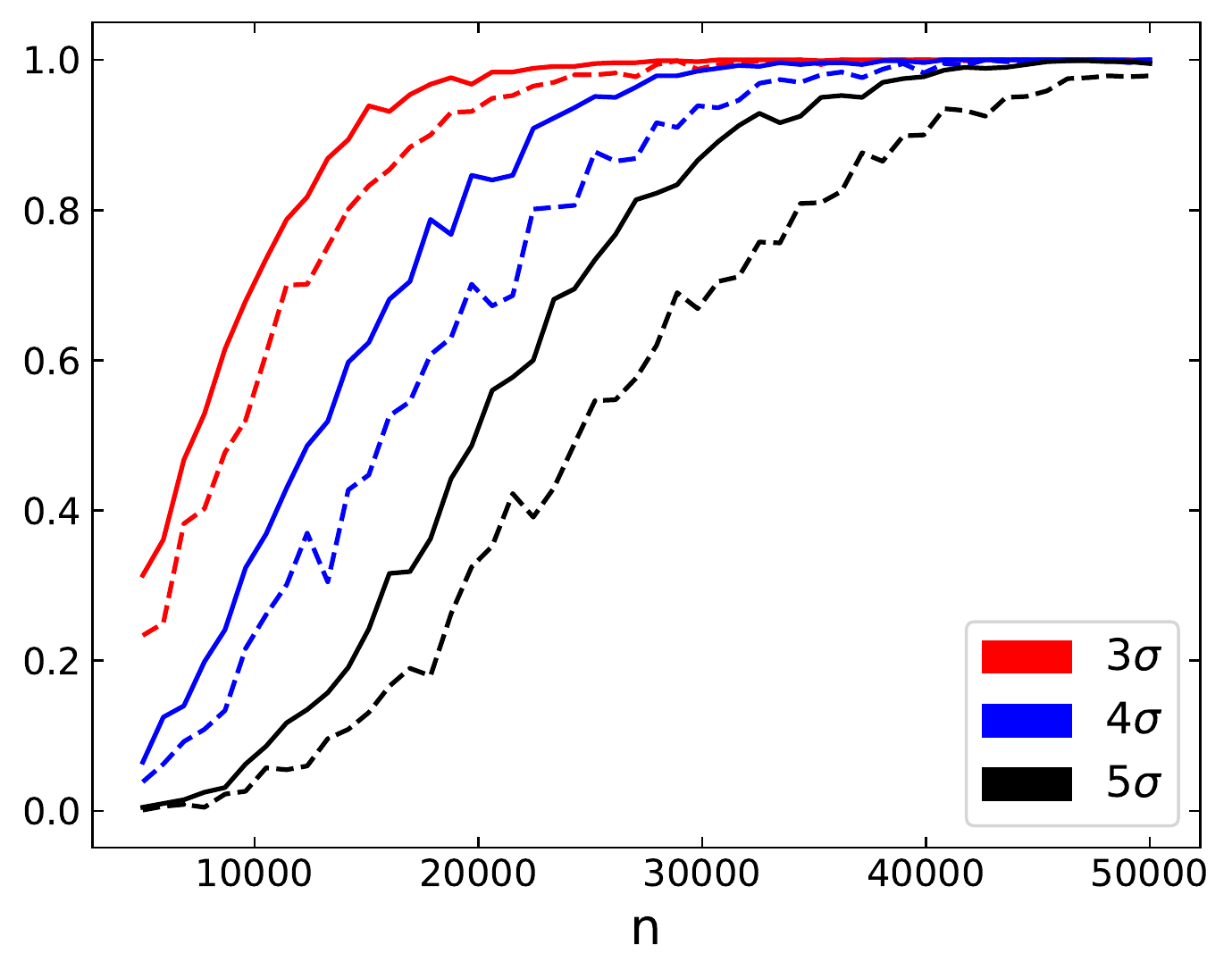}}
	\caption{(\textit{Top}) Statistical powers for the PRS (\textit{solid lines}) and the HRO shape parameter (\textit{dashed lines}) for $N=1000$ samples of $n=5000$ angles from the von Mises distribution $\mathcal{F}(\theta \vert \mu,\kappa)$, with thresholds of positive detection $3,4$ and $5\sigma$ for varying $\kappa$. (\textit{Bottom}) Statistical powers for the PRS and HRO shape parameter for fixed $\kappa = 0.05$ and thresholds $3,4$ and $5\sigma$ for $N = 1000$ samples of $n$ angles, with $n$ varying from 5000 to 50,000.}
	\label{fig:power}
\end{figure}

Since, for small $\kappa$, the von Mises distribution approaches a uniform distribution, we can vary $\kappa$ in order to examine the relative power of the statistics. The top panel of Fig.~\ref{fig:power} shows the statistical power of both statistics for varying $\kappa$, for thresholds of $3$, $4$ and $5\sigma$. Both powers converge to 0 as the distribution becomes increasingly uniform, i.e. for small $\kappa$, while as $\kappa$ increases both powers increase to the maximum, 1. However, the figure shows that the power of the PRS increases more quickly to 1, and decreases more slowly to 0. The largest difference between the two statistics occurs for intermediate values of $1/\kappa$, with the PRS accurately detecting non-uniformity 24 per cent more of the time. The bottom panel of Fig.~\ref{fig:power} shows the statistical powers calculated in the same way, but fixing $\kappa$ to be 0.05, and allowing the number of sample angles to increase. The value $\kappa = 0.05$ characterises a von Mises distribution that is very close to uniform, and so a large sample size is required for both statistics to consistently pick out its non-uniformity. As the sample size increases, we find that the sensitivity of the PRS increases more quickly than the HRO shape parameter, showing that it is more sensitive at picking out orientation effects. For the case of a $4\sigma$ detection threshold, the statistical power of the PRS first comes to within 0.01 of 1 for samples of size 33000, as opposed to 39000 for the HRO statistic.

We have shown that for a sufficiently large number of angles the PRS and the HRO shape parameter follow normal distributions, for a simple model of uniformly distributed relative angles, and for two uncorrelated pseudo-vector fields. As a test against the uniformity of the relative angles for the simulated data, the PRS has greater statistical power than the HRO shape parameter. We now apply and compare the two statistics for the BLASTPol polarization and \textit{Herschel} derived column density observations of the Vela C molecular cloud.

\section{Application to Sky Polarization}
\label{sec:polarization}

\citet{2017A&A...603A..64S} have applied the HRO method to study the relative orientations of the magnetic field inferred from BLASTPol observations and the iso-column density contours (orthogonal to the gradient of the column density) inferred from Herschel observations of the Vela C Molecular Cloud (see Section 2).  For comparison, we apply the PRS and the HRO statistic to the same data.

Given that the resolution of the BLASTPol observations (3.0\,arcmin FWHM) is smaller than the resolution of the \textit{Herschel} observations (35.2\,arcsec FWHM), we are comparing large-scale orientations with smaller-scale cloud features traced by the column density. One could also simply use the BLASTPol observations for Stokes $I$ as a proxy for column density in order to have both polarization and column density data at the same scale. However, we prefer to use the higher resolution \textit{Herschel} data since this gives a better estimate of the gradient of the column density. We also use the \textit{Herschel}-derived column density for consistency with \citet{2017A&A...603A..64S}. 

\subsection{Calculating the relative angles}
\label{sec:Pol_phi}

With the polarization pseudo-vector  $\bf{P}_\lambda$ taken to be perpendicular to the magnetic field $\langle\hat{\bf{B}}_{\perp}\rangle$, we can calculate the angle, $\phi$, between the magnetic field and the tangent to the $N_\mathrm{H}$ contours as

\begin{equation}
	\phi_\lambda = \arctan\frac{ \lvert \nabla N_\mathrm{H} \times \bf{P}_\lambda \rvert}{ \nabla N_\mathrm{H} \cdot \bf{P}_\lambda},
	\label{eq:Nh_pol_relative_angle}
\end{equation}
where, as implemented, the norm carries a sign \citep{2017A&A...603A..64S}. The gradient of the column density was computed using a $5 \times 5$ Sobel kernel.

We mask pixels where the polarization signal-to-noise ratio is less than 3, where the polarization intensity is defined by $P_\lambda \equiv \sqrt{U_\lambda^2 + Q_\lambda^2}$. This technique has previously been applied to analyses involving the polarization angle \citep[e.g.,][]{2012ApJS..201...13V,2017A&A...603A..64S} and has been shown to reduce the uncertainty in polarization angle to negligible \citep{1958AcA.....8..135S,1993A&A...274..968N}.

Note that one should also ensure that the sample angles are independent (i.e. that the individual pixels are not small compared with the beamsize); however, we keep all the pixels for consistency with \citet{2017A&A...603A..64S}. For the comparison of the BLASTPol and \textit{Herschel} observations, each gradient vector provides an independent measurement of the iso-NH contours, therefore the statistical significance is given by those observations, independent of the scale at which we sample the magnetic field. Thus, we do not expect a significant difference between analyses performed on data with and without a sampling correction. Moreover, since we are only comparing the relative performance of the HRO and PRS approaches there is no benefit to the added complexity.

\subsection{Comparing statistics in the Vela C molecular cloud}
\label{sec:VelaC_comparison}

\begin{figure}
\includegraphics[width=\columnwidth]{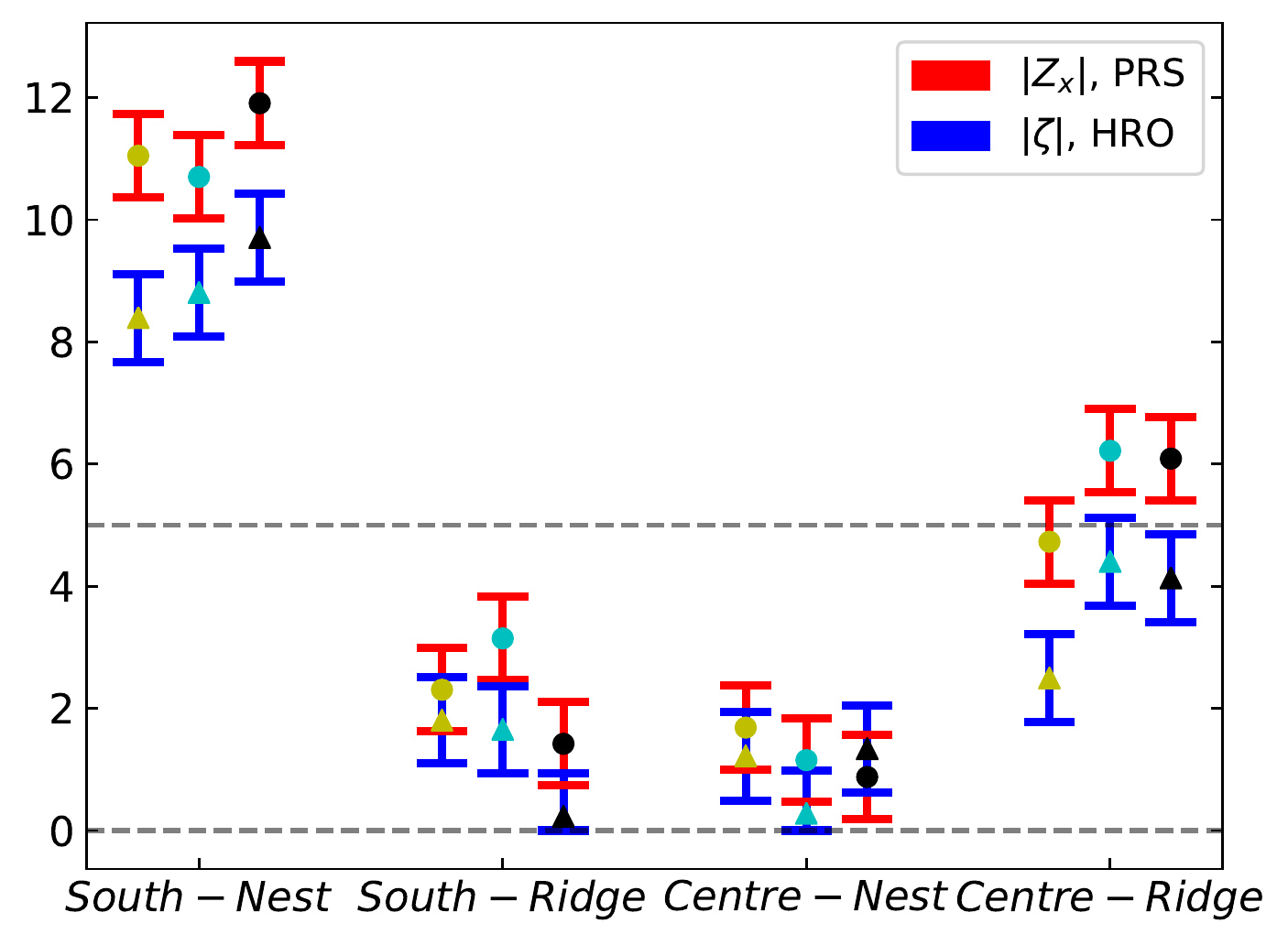}
	\caption{Comparison of the statistical significance of the absolute value of the PRS, $|Z_x|$, and the HRO shape parameter, $|\zeta|$, for four rectangular regions centred around the South-Nest, South-Ridge, Centre-Nest, and Centre-Ridge of Vela C defined in \citet{2011A&amp;A...533A..94H}, for each of the BLASTPol 250-, 350-, and 500-$\mu$m observations. The statistical significance, on the y axis, is simply the signal given in units of the relevant $\sigma$; in these units, the error bars correspond to uncertainties of $\pm 1 \sigma$. The red bars with circle markers correspond to the PRS, while the blue bars with triangle markers correspond to the HRO statistic. The yellow, green, and black marker colours correspond to the 250-, 350- and 500-$\mu$m observations, respectively. The number of relative pixels available in the South-Nest,  South-Ridge, Centre-Nest, and Centre-Ridge is $n$ = 45000, 31500, 21000, and 50625, respectively. The horizontal grey dashed lines correspond to 0$\sigma$ and a common conservative detection limit of 5$\sigma$. The greatest increase in significance occurs for the South-Nest region in the 250-$\mu$m band, with the PRS having a greater significance by a factor of 1.3}
	\label{fig:ZHRO_VelaC}
\end{figure}

The overlap between the $N_\mathrm{H}$ map derived from the \textit{Herschel} observations and the BLASTPol observations used in this study contains four of the sub-regions defined by \citet{2011A&amp;A...533A..94H}, as can be seen in Fig.~\ref{fig:VelaC}. In order to compare the statistical power of the PRS and the HRO shape parameter as applied to polarization data, we compute them separately for rectangular regions centred around each of these sub-regions, and for the three different submillimetre bands. The results are summarized in Fig.~\ref{fig:ZHRO_VelaC}. 


We find that, as anticipated, the PRS tends to produce a stronger signal for the same amount of alignment between the column density gradient and the polarization pseudo-vector when compared with the HRO shape parameter. In particular, the PRS results have higher statistical significance when the alignment is greater. The difference between the two statistics in all three BLASTPol wavebands is largest for the South-Nest and Centre-Ridge regions, for which the greatest number of relative angles exist. The greatest increase in significance occurs for the South-Nest region in the 250-$\mu$m band, with the PRS resulting in 31 per cent greater significance. In addition to the PRS being more sensitive to alignment between the column density gradient and polarization in general, we further note that its sensitivity increases as the number of pixels with useful polarization data increases.

\begin{figure*}
\subfigure{\includegraphics[width=\columnwidth]{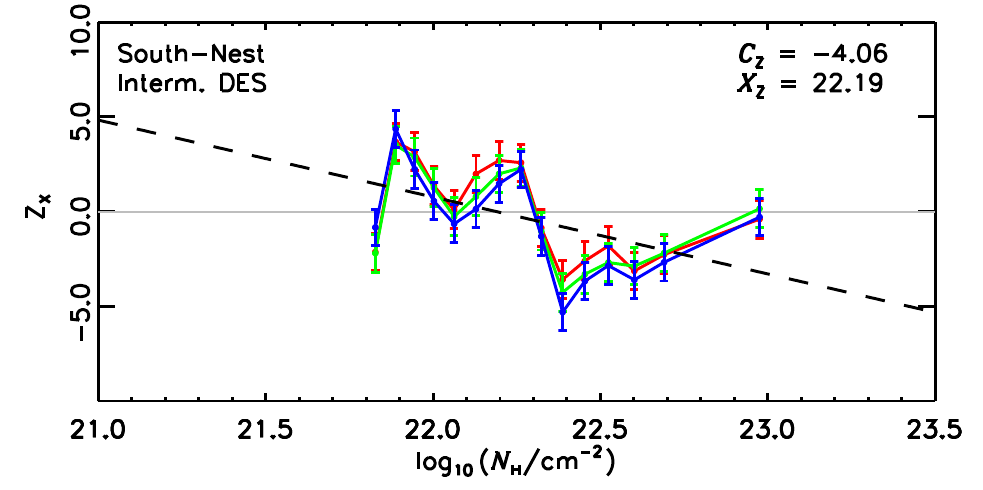}}
\subfigure{\includegraphics[width=\columnwidth]{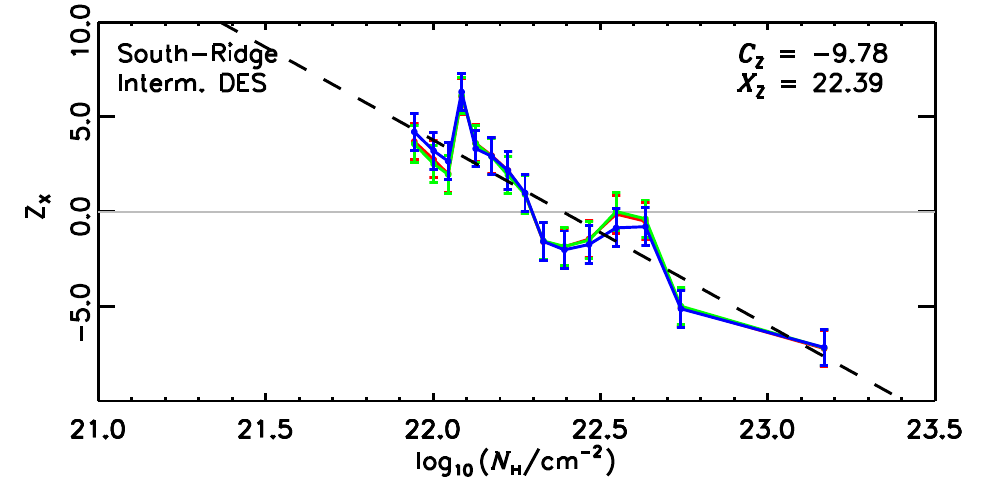}}
\subfigure{\includegraphics[width=\columnwidth]{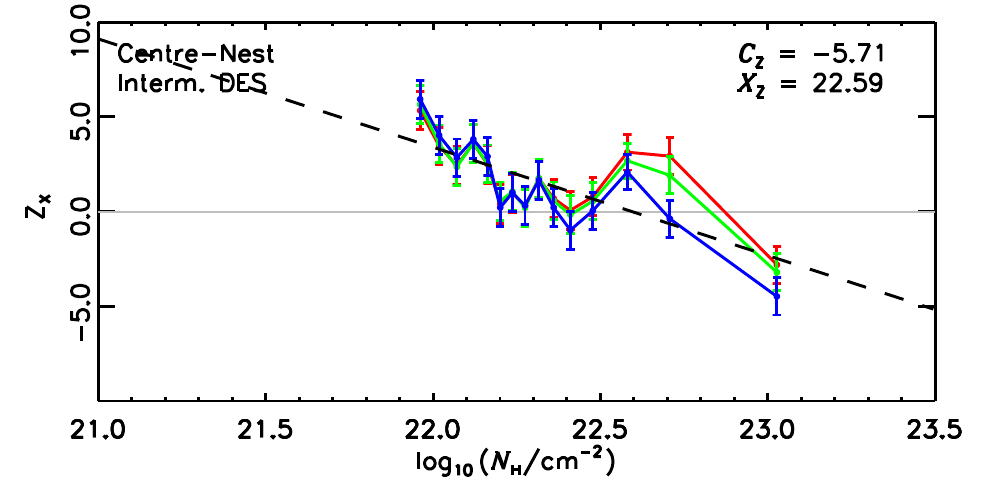}}
\subfigure{\includegraphics[width=\columnwidth]{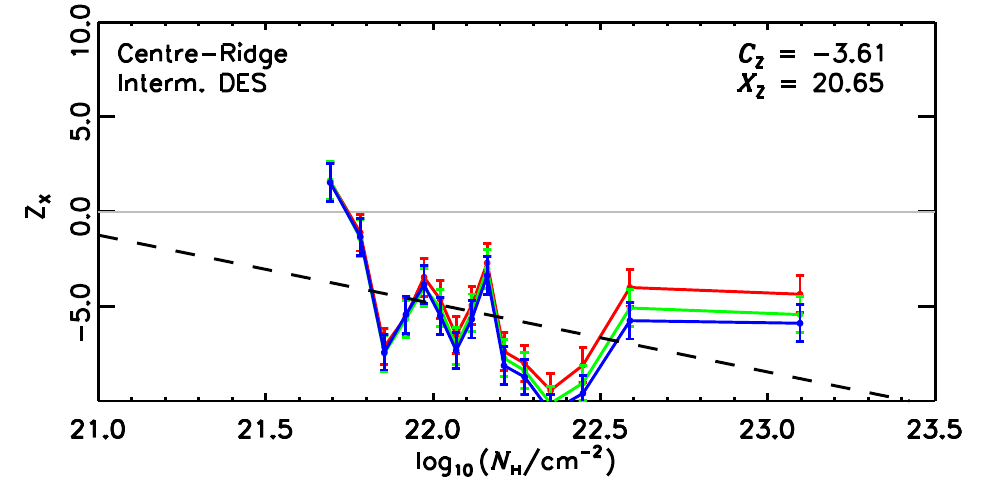}}
	\caption{PRS, $Z_x$, calculated for different $N_\mathrm{H}$ bins. The black dashed line and the values of $C_Z$ and $X_Z$ correspond to the linear fit $Z_x = C_Z[\log_{10}(N_\mathrm{H}/\mathrm{cm}^{-2})-X_Z]$. The grey line is $Z_x = 0$. Separate panels show results for four sub-regions of the Vela C molecular cloud as defined in \citet{2011A&amp;A...533A..94H} and shown in Fig.~\ref{fig:VelaC}.}
	\label{fig:Soler_reproduction}
\end{figure*}

We also examine the relative orientations for varying column densities for each of the four sub-regions of the Vela C molecular clouds \citep[as in][]{2017A&A...603A..64S}. For each of the four sub-regions, we bin the column density into 15 bins, such that each bin has the same number of relative angles available. For each bin, we then calculate the PRS. Figure~\ref{fig:Soler_reproduction} shows the results of these calculations, and can be compared to Fig.~9 in \cite{2017A&A...603A..64S}, which does the same calculations with the HRO shape parameter. Using the PRS, in all four sub-regions we confirm that with increasing column density the iso-column density contours become increasingly perpendicular to the magnetic field.There is a value of the column density at which both the PRS and HRO shape parameter are close to zero, and the alignment switches from preferentially parallel to perpendicular. We find that the values of column density at which the alignment switches from preferentially parallel to perpendicular recovered using the PRS and the HRO shape parameter are consistent. Furthermore, when using the PRS the uncertainties are reduced, so that in regions of column density where the relative orientations are increasingly uniform the PRS can still detect whether the orientations are preferentially parallel or perpendicular. This is expected, given that the uncertainties in \cite{2017A&A...603A..64S} correspond to those estimated from the variance in the histogram counts and therefore, overestimate the dispersion with respect to an homogeneous distribution of angles, which is precisely what the $Z_{x}$ uncertainties are representing. In Appendix \ref{app:GouldBeltClouds}, we use the PRS to re-analyse the relative orientation between the $N_\mathrm{H}$ structures and magnetic field, inferred from the \textit{Planck} 353-GHz polarization observations, towards 10 Gould Belt molecular clouds, which was originally carried out using the HRO shape parameter in \citet{2016A&A...586A.138P}. 

The physical conditions responsible for the observed change in relative orientation between the column density and polarization (taken here to characterise the direction of the magnetic field projected on the plane of the sky) are related to the degree of magnetisation of the molecular clouds \citep{2013A&A...556A.153H,2013ApJ...774..128S}. These trends in relative orientation between column density structures and the magnetic field have been examined via simulations of molecular clouds \citep[e.g.,][]{2013ApJ...774..128S,2016ApJ...829...84C} and have been found to be in agreement with the classical picture of molecular cloud formation, in which the cloud forms following the compression of background gas by the passage of the Galactic spiral shock or an expanding supernova shell, and the compressed gas cools flowing down the magnetic field lines to form a self-gravitating mass \citep{1965QJRAS...6..161M,1984A&A...136...98M}. More detailed discussion of the physical significance of these results can be found in \citet{2017A&A...603A..64S}.

In this work, we use the PRS to test whether the relative orientation is preferentially parallel or perpendicular. We note, however, that a measurement of $Z_x$ would not be relevant for testing for a preferred relative orientation of $45^\circ$. As discussed elsewhere \citep[e.g.,][]{10.2307/30080924,ANESHANSLEY:1981aa}, it is possible to check whether the orientation prefers some non-trivial angles (i.e. that is not along the $x$-axis) by checking the projection along the $y$-axis, i.e. $Z_y$. We leave a comparison of $Z_x$ and $Z_y$, and considerations on the relative orientation from 3-dimensional vectors projected on to the plane of the sky for a future study.
 
\subsection{Polarization bias}
\label{sec:Polarization_bias}

We can improve the application of the PRS to polarization data by being slightly more sophisticated in our approach to the uncertainty in polarization angle. Previously, we reduced the effect by considering only the angles calculated for pixels in which the polarization intensity signal-to-noise ratio (S/N), $P/\sigma_P$, was greater than 3. \citet{2015A&A...574A.136M} suggest, however, that this choice of threshold might not be appropriate for every measurement, and, in any case, we would still like to utilise even low S/N data in order to access the full power of the data. 

To do this, we use Eq.~\ref{eq:Zx_weighted} for the weighted PRS, adopting the uncertainty in $P$ in place of the angle uncertainty. In principle, it would be better to use the variance in polarization angle as the weight. However, in practice this is a non-linear, but monotonic, function of the polarization \citep[see][]{ 2015A&A...574A.136M}. Hence, there is no real benefit in adding complexity, and we simply use $P/\sigma_P$ as a proxy. Here we are also assuming that the variance in the $N_\mathrm{H}$ gradient angle is small compared to the variance in the polarization angle, so that the noise in the relative angles is dominated by the polarization noise. For more general applications of the PRS this will not be true, and noise from both sets of orientations being compared must contribute to the weighting.

\begin{figure}
	\includegraphics[width=\columnwidth]{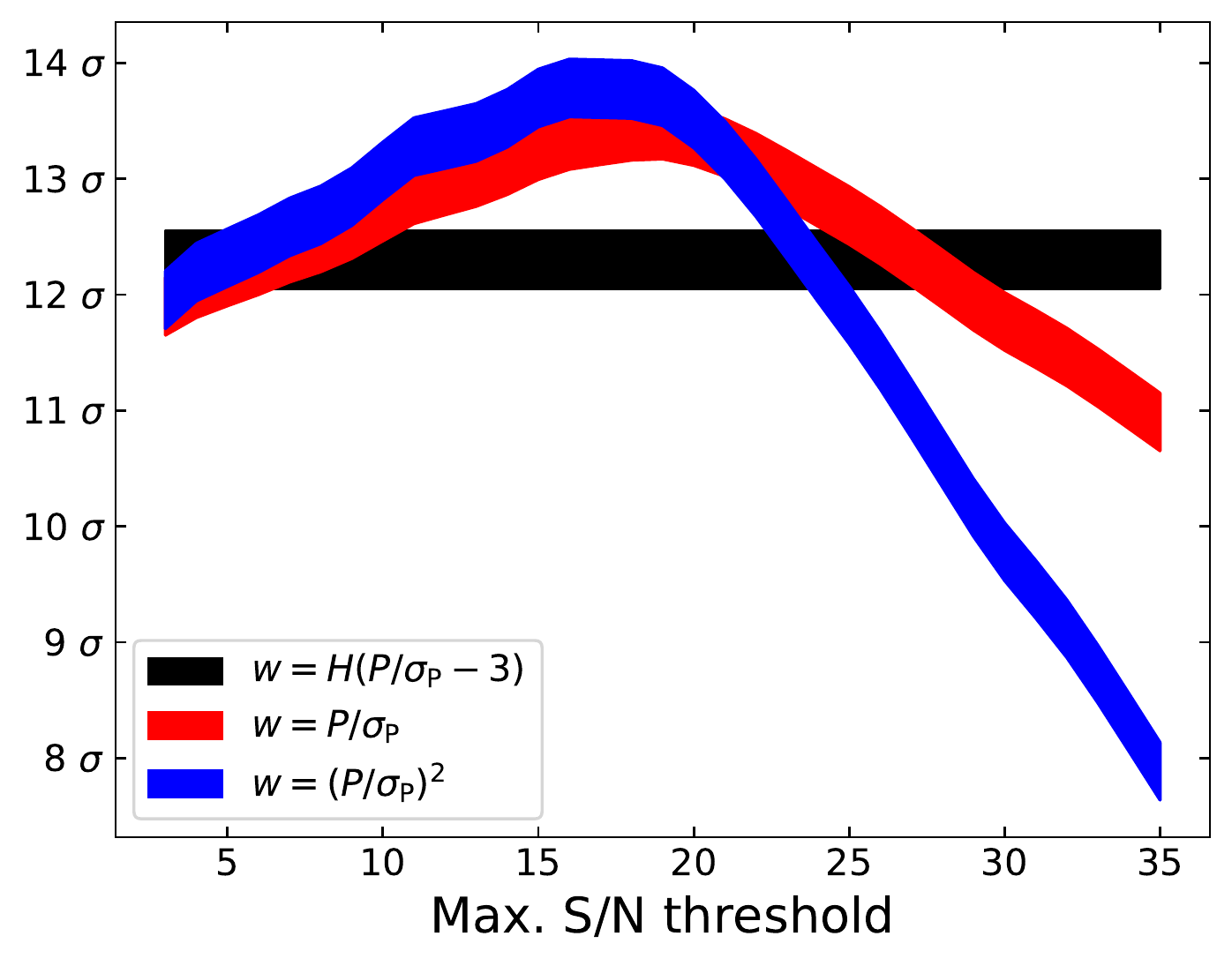}
	\caption{Statistical significance of the PRS, when weighted according to different weighting schemes based on the polarization intensity S/N, as a function of maximum S/N threshold beyond which all data are weighted the same. The maximum S/N threshold corresponds to the S/N beyond which all angles are weighted the same. Here $H(x)$ is the Heaviside stepfunction. The widths of the lines correspond to the $1\sigma$ uncertainty. The results are computed for the entire Vela C region contained in the overlap between the BLASTPol and \textit{Herschel} maps (shown in Fig.~\ref{fig:VelaC}), which includes a total of 325000 pixels.}
	\label{fig:SNR_cutoffs}
\end{figure}

We calculate the relative angles for every pixel, regardless of their polarization intensity S/N, and then calculate the PRS applying lower weights in Equation 8 to lower S/N data. A maximum S/N threshold must be chosen above which all angles are weighted the same, since otherwise a small number of points with very high S/N will dominate. Figure~\ref{fig:SNR_cutoffs} shows the weighted PRS calculated for the entire Vela C molecular cloud for various choices of this maximum S/N threshold. Three different weighting schemes were chosen: the weights (up to the threshold) were taken to be either $w=P/\sigma_P$,  $w=(P/\sigma_P)^2$, or they were taken to be 0 for all pixels in which S/N$\,{<}\,3$ and 1 otherwise (i.e. $w=H(P/\sigma_{P}-3)$, where $H(x)$ is the Heaviside step function). This latter scheme is equivalent to our previous method of simply ignoring pixels in which the S/N was less than 3.

Figure~\ref{fig:SNR_cutoffs} shows the PRS calculated for different maximum S/N thresholds for the entire Vela C region (shown in Fig.~\ref{fig:VelaC}). There is a significant increase in the reported alignment when all of the relative angles are being utilised and weighted according to the polarization intensity S/N, compared to when data in which the S/N<3 are rejected. The largest improvement occurs when the maximum S/N threshold is approximately 18, and for the scheme in which the weights are taken to be the S/N squared, with an improvement of approximately 10 per cent. Beyond this threshold the improvement is rapidly lost as the crucial information is lost in the weight coming from the few high S/N pixels.  

In order to utilise all polarization data available, we suggest the use of a scheme in which the terms of the PRS are weighted according to the square of the polarization intensity S/N, up to some maximum threshold. We find that a threshold of S/N=18 is best for this particular data set; however, the appropriate choice of threshold will, in general, depend on the S/N distribution and should be determined for each analysis individually.

We have focused our analysis on the statistical errors in the polarization data. However, in the Vela C observations, the larger uncertainty in polarization angle is due to the uncertainty in separating the diffuse polarized emission from dust in the foregrounds and backgrounds from the polarized dust emission from the Vela C molecular cloud. Further investigation should be conducted into the effects on the PRS values of the different diffuse ISM subtraction methods presented in \citet{2016ApJ...824..134F}, as was done for the HRO shape parameter in \citet{2017A&A...603A..64S}. This would help to understand the amplitude of the systematic uncertainties due to component separation.

\section{Conclusions}

We have presented a robust and efficient method for determining whether two pseudo-vector fields tend to align, either preferentially perpendicular, or preferentially parallel. Our approach uses the projected Rayleigh statistic, which performs better than binning statistics that have previously been used. 

We find that as a test against the uniformity of the distribution of relative angles the PRS is more statistically powerful, and, in particular, loses sensitivity slower than the HRO shape parameter as the distribution of angles approaches uniformity. The sensitivity of the PRS (compared to the HRO shape parameter) also improves faster with larger sample sizes. 

This study was motivated by the astrophysical problem of characterising the relative alignment of the magnetic field projected on the plane of the sky and the orientations of structures in the interstellar medium. When applying the two statistics to the magnetic field inferred from BLASTPol polarization data and column density inferred from \textit{Herschel} observations, for various sub-regions of the Vela C molecular complex, we find that, indeed, the projected Rayleigh statistic reports a higher statistical significance for the correlations. Using the polarization intensity signal-to-noise ratio as a proxy for the polarization angle noise in a weighting scheme, we find that we can further improve the significance. 

In addition to what we have already discussed, one might also be interested in examining the scale over which the correlation in orientation occurs. For example, one could use either or both of the PRS or the HRO shape parameter to construct a cross-correlation function for different displacements, with what we have measured here being the zero lag value of this function. The full cross-correlation function could then be used to determine the scale at which the pseudo-vector fields become effectively de-correlated. However, this is somewhat complicated by the fact that the column density structures and polarization fields themselves have auto-correlations. Thus any analysis of scale must somehow distinguish between the effects of structures in each field individually, and of actual de-correlation between the two fields. This was not attempted in this work, but may be a direction for future investigation.

Beyond magnetic fields and the interstellar medium, many astronomical phenomena include orientation effects. The method presented here may be applied to analysing these effects, by examining preferences between various types of orientation data. For example, the relative orientations of individual galaxies within larger scale structures may be characterised with this method. Binary systems also have orientations characterised by their orbital planes, which may be aligned with larger scale orientations as well. Thus, the method presented here has the potential for a broad range of applications in astronomy.

\section*{Acknowledgements}

\textit{Herschel} is an ESA space observatory with science instruments provided by European-led Principal Investigator consortia and with important participation from NASA. The BLASTPol collaboration acknowledges support from NASA through grant numbers NAG5-12785, NAG5-13301, NNGO-6GI11G, NNX0-9AB98G, and the Illinois Space Grant Consortium, the Canadian Space Agency, the Leverhulme Trust through the Research Project Grant F/00 407/BN, Canada's Natural Sciences and Engineering Research Council, the Canada Foundation for Innovation, the Ontario Innovation Trust, and the US National Science Foundation Office of Polar Programs. LMF is a Jansky Fellow of the National Radio Astronomy Observatory (NRAO). NRAO is a facility of the National Science Foundation (NSF) operated under cooperative agreement by Associated Universities, Inc. FP is a Marie Sklodowska-Curie fellow from the European Union's Horizon 2020 research and innovation programme under grant agreement No 658499.This work was possible through funding from the Natural Sciences and Engineering Research Council of Canada.





\bibliographystyle{mnras}
\bibliography{Bibfile} 



\appendix

\section{Reanalyzing the Gould Belt clouds}\label{app:GouldBeltClouds}

For the sake of comparison, we repeat the analysis of the relative orientation between the $N_\mathrm{H}$-structures and the orientation of the magnetic orientation, inferred from the {\it Planck} 353-GHz polarization observations, towards 10 Gould Belt molecular clouds, originally presented in \citet{2016A&A...586A.138P}.
We use the same gradients of $N_\mathrm{H}$ and polarization observations selected for the \citet{2016A&A...586A.138P} analysis, but we replaced the calculation of the histograms of relative orientation and the histogram shape parameter, $\zeta$, for the projected Rayleigh statistic, $Z_{x}$.

\begin{figure*}[ht!]
\centerline{
\includegraphics[width=0.5\textwidth,angle=0,origin=c]{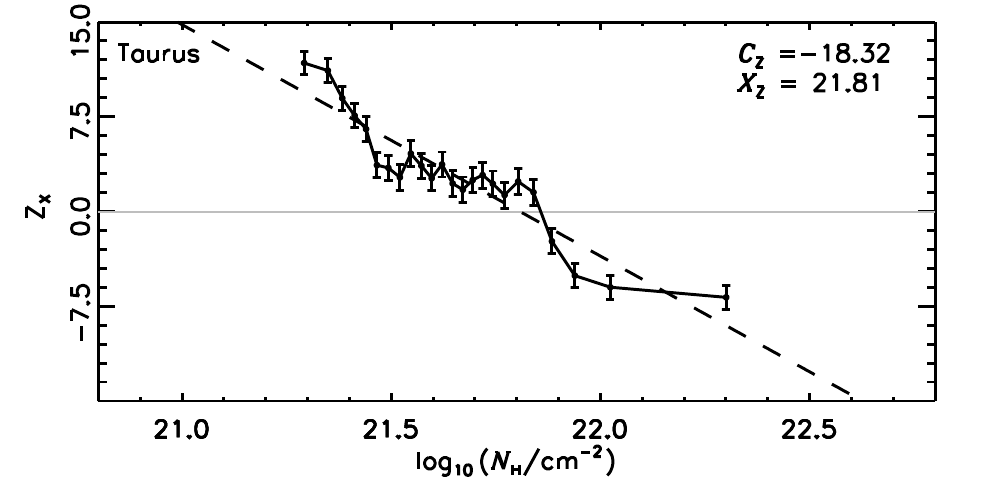}
\includegraphics[width=0.5\textwidth,angle=0,origin=c]{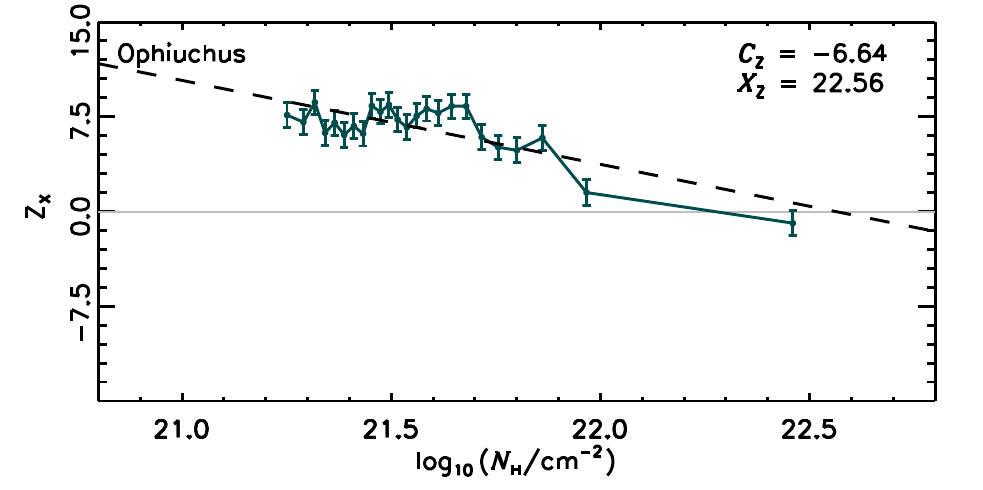}
}
\vspace{-2.0mm}
\centerline{
\includegraphics[width=0.5\textwidth,angle=0,origin=c]{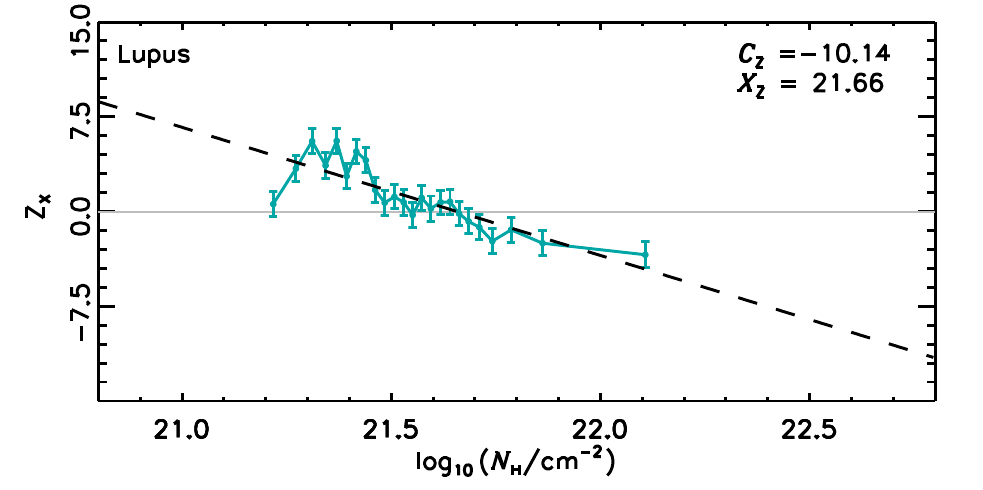}
\includegraphics[width=0.5\textwidth,angle=0,origin=c]{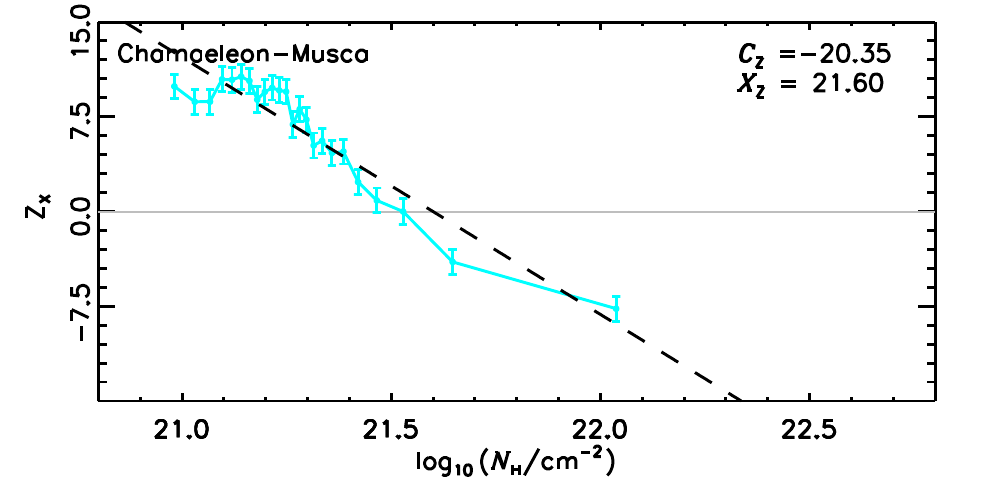}
}
\vspace{-2.0mm}
\centerline{
\includegraphics[width=0.5\textwidth,angle=0,origin=c]{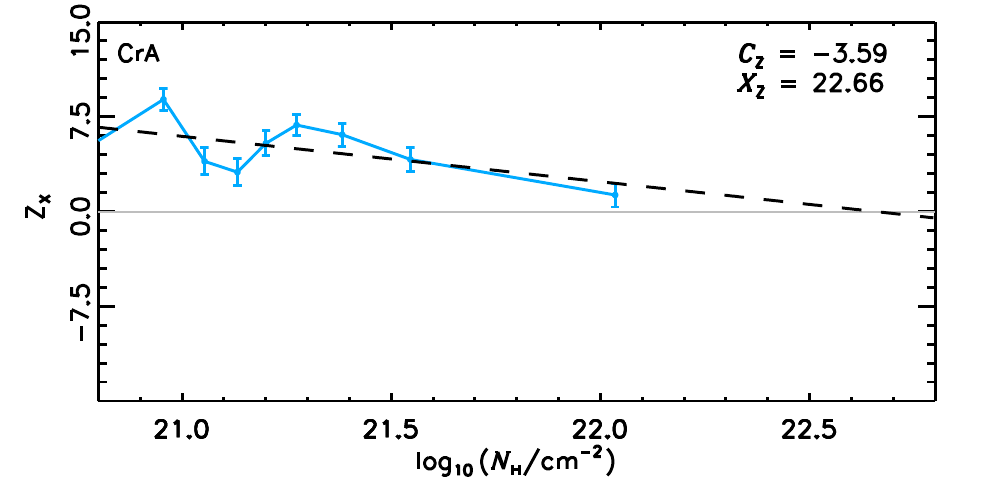}
\includegraphics[width=0.5\textwidth,angle=0,origin=c]{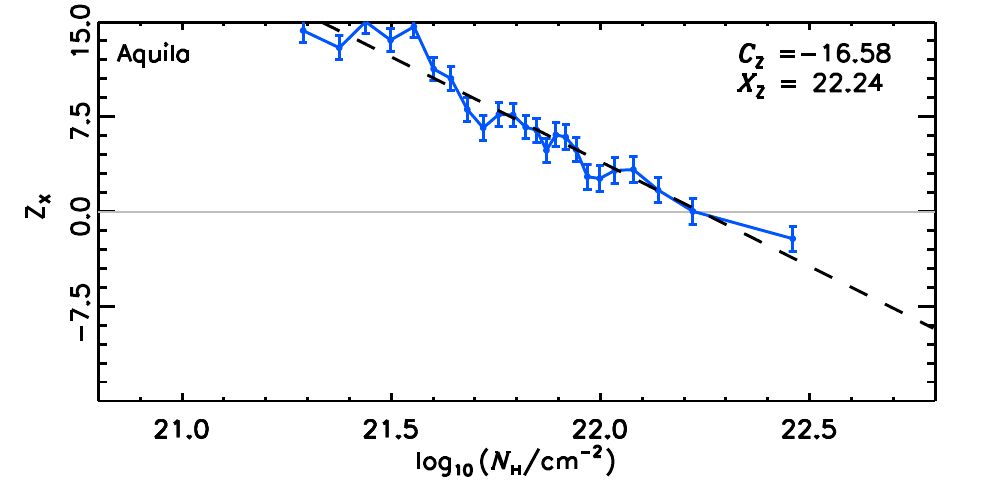}
}
\vspace{-2.0mm}
\centerline{
\includegraphics[width=0.5\textwidth,angle=0,origin=c]{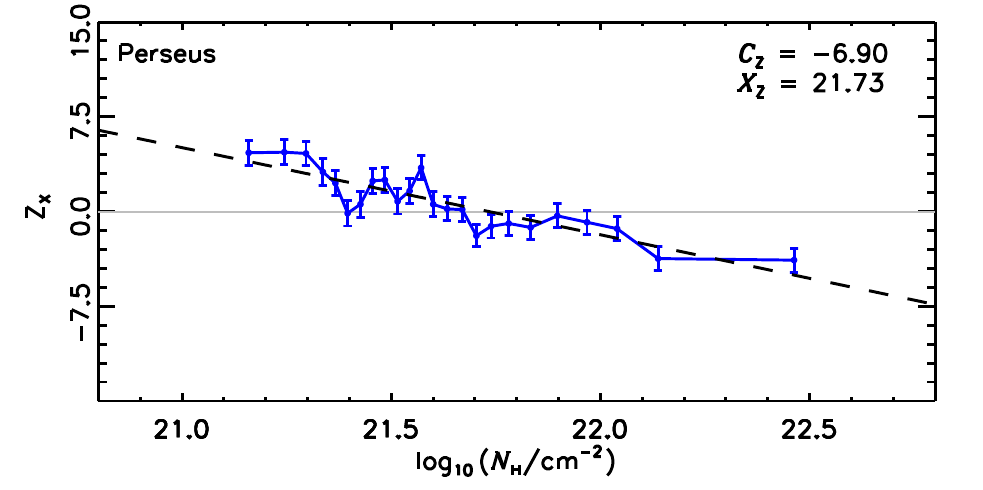}
\includegraphics[width=0.5\textwidth,angle=0,origin=c]{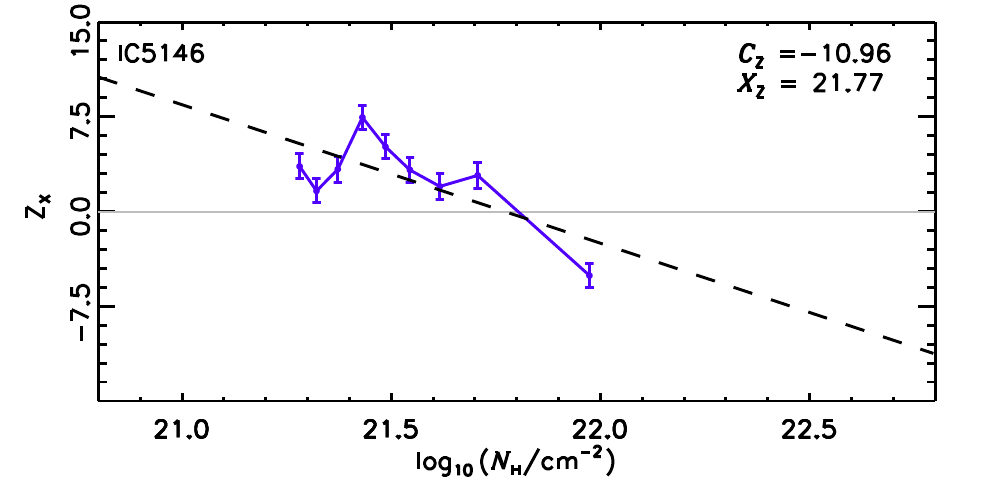}
}
\vspace{-2.0mm}
\centerline{
\includegraphics[width=0.5\textwidth,angle=0,origin=c]{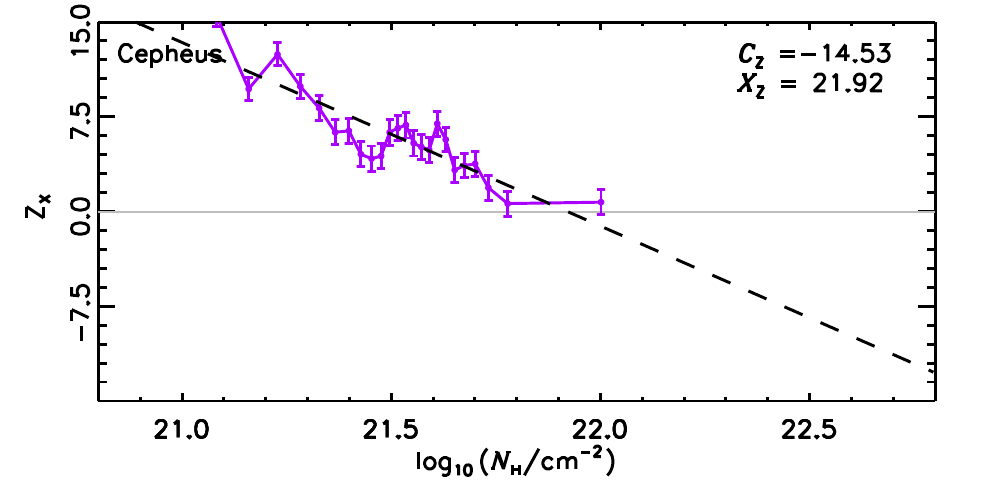}
\includegraphics[width=0.5\textwidth,angle=0,origin=c]{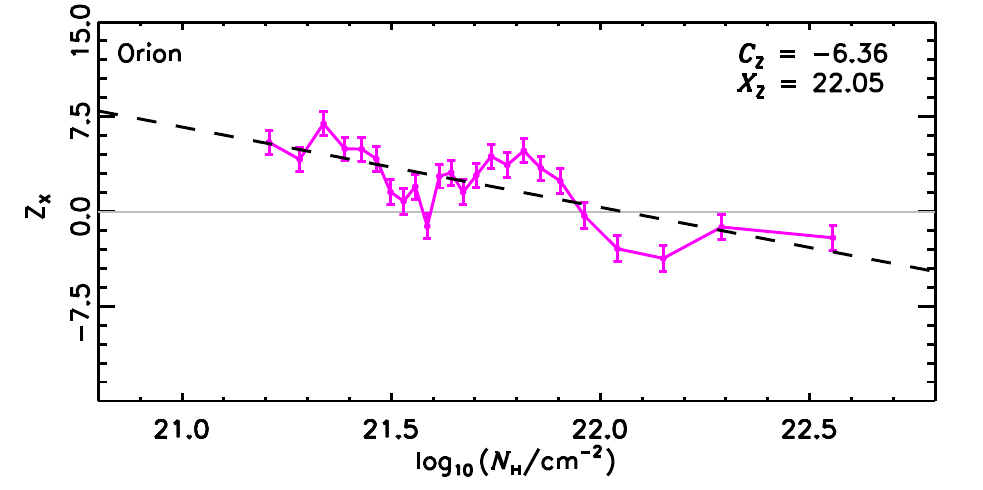}
}
\vspace{-1.0mm}
\caption{Rayleigh statistic.}\label{fig:PRSmulti}
\end{figure*}

The values of $Z_{x}$, presented in Figure~\ref{fig:PRSmulti}, show that the trends in relative orientation presented in terms of  $\zeta$ in figure~7 of \citet{2016A&A...586A.138P} are also recovered using the PRS.
The error bars in $Z_{x}$ are significantly smaller that those in $\zeta$.
This is as expected, for the same reasons discussed in Sec.~\ref{sec:VelaC_comparison}.

Figure~\ref{fig:PRSmulti} shows that the relative orientation between the $N_\mathrm{H}$-structures and the magnetic field orientation in the highest $N_\mathrm{H}$-bins in Ophiuchus, Lupus, Aquila, and Orion is significantly more perpendicular than what could be inferred from the values of $\zeta$.
However, there is indication of a preferential relative orientation in the highest $N_\mathrm{H}$-bins towards CrA and Cepheus, both in terms of $\zeta$ or $Z_{x}$.
The decrease in the values of $Z_{x}$ with increasing $N_\mathrm{H}$ is consistent with that reported in \citet{2016A&A...586A.138P}, and the values of $N_\mathrm{H}$ where the relative orientation changes from mostly parallel to mostly perpendicular, $X_{\rm HRO}$ in \citet{2016A&A...586A.138P} or  $X_{\rm Z}$ in this study, are remarkably close in both analyses.

\bsp	
\label{lastpage}
\end{document}